\documentclass[prc,aps,showpacs]{revtex4}
\usepackage{graphicx}
\usepackage{amsmath}

\def\pcomma{$^,$}
\def\pgwu{$^1$}
\def\pduke{$^2$}
\def\pjlab{$^3$}
\def\ppnpi{$^4$}
\def\puva{$^5$}

\begin{document}

\vspace*{0.75in}
\begin{center}
{\Large\bf Properties of the Resonance $\Lambda(1520)$ as seen in 
	the Forward Electroproduction at JLab Hall A \\}
\vspace*{0.1in}
{\large I.I.~Strakovsky\pgwu,
	Y.~Qiang\pduke\pcomma\pjlab,
        Ya.I.~Azimov\ppnpi,
        W.~J.~Briscoe\pgwu,
        H.~Gao\pduke,
        D.W.~Higinbotham\pjlab, and
        V.V.~Nelyubin\puva \\}
\vspace*{0.1in}
{\small\it\pgwu The George Washington University, Washington, DC
        20052, USA \\}
{\small\it\pduke Duke University Triangle Universities Nuclear Laboratory,
            Durham, NC 27708, USA \\}
{\small\it\pjlab Thomas Jefferson National Acceleration Facility,
            Newport News, VA 23606, USA \\}
{\small\it\ppnpi Petersburg Nuclear Physics Institute, 188300
        Gatchina, Russia \\}
{\small\it\puva University of Virginia, Charlottesville, VA 22901, USA \\}
\end{center}

\textbf{Abstract.} {\small High-resolution spectrometer 
measurements of the reaction $H(e,e^\prime K^+)X$ at 
small $Q^2$ are used to extract the mass and width of 
the $\Lambda(1520)$. We investigate dependence of the 
resonance parameters on different parametrizations of 
the background and the resonance peak itself. Our 
final values for the Breit-Wigner parameters are 
$M=1520.4\pm0.6$(stat.)$\pm1.5$(syst.)~MeV and
$\Gamma=18.6\pm1.9$(stat.)$\pm1$(syst.)~MeV. 
The width appears to be more sensitive to the 
assumptions than the mass. We also estimate, for 
the first time, the pole position for this resonance 
and find that both the pole mass and width seem to 
be smaller than their Breit-Wigner values.}

\section{Introduction}

The $\Lambda(1520)$ is considered to be one of the best 
known baryon resonances. It is an excited hyperon state 
of the highest rank (4 stars) in the Review of Particle 
Physics (RPP)~\cite{PDG}. But its properties are still 
not exceedingly well understood. For example, its rather 
intensive decay to the $\pi\pi\Lambda$ channel~\cite{PDG} 
gives a hint at an unexpectedly strong coupling to the 
kinematically suppressed subchannel $\pi\Sigma(1385)$, 
stronger than to the main decay channels $\pi\Sigma$ and 
$\overline{K}N$. This could mean that the $\Lambda(1520)$ 
has a molecular nature (see, e.g., Ref.~\cite{oset} and 
references therein). On the other hand, evidence has been
suggested~\cite{zou} that a new, previously unobserved, 
resonance $\Sigma(1380)$ with $J^P=1/2^-$ and decay to 
$\Lambda\pi$ might influence the mode $\Lambda(1520)\to\pi
\pi\Lambda$. These examples show that more detailed studies 
of the $\Lambda(1520)$, which may clarify both its nature and 
the hyperon spectroscopy, are needed. 

As a first step, we investigate degree of precision available 
for the basic resonance parameters of the 
$\Lambda(1520)$~\cite{qi10}. Surprisingly, all the experimental 
inputs shown in RPP~\cite{PDG} for its mass and width are dated 
before 1980. Many new experiments performed since then have not 
been accounted for. Meanwhile, they provided much more data with 
the $\Lambda(1520)$ observed either as the main goal or as a 
byproduct (even more data are expected to appear in near future). 

Most of the earlier high-precision mass and width values, 
in particular those used for averaging in RPP~\cite{PDG}, had 
come either directly, from production measurements with bubble 
chambers, or indirectly, from partial wave analyses of scattering
data. High-energy spectrometers did not have good enough 
resolution for such measurements, and their results could not 
compete with bubble chamber data, even despite much higher 
statistics.  Now, the High Resolution Spectrometers 
(HRS's)~\cite{HRS} in the JLab Hall~A experiment~\cite{qi07}
provided an exciting accuracy ($\sigma=1.5$~MeV), comparable 
to the best previous resolutions. That is why their measurements 
are competitive with any previous ones.

Note also that experiments used for averaging in RPP, applied very 
different assumptions to extract the resonance parameters. However,
influence of those assumptions on the parameter values has never
been analyzed. We investigate this problem~\cite{qi10} on the base 
of the Hall~A data set~\cite{qi07}.

\section{ JLab Hall~A Experiment}

High-resolution measurements of the missing mass (MM) spectra
were developed at near-forward production angles in the 
reactions $ep\to e^\prime K^\pm X$ and $e^\prime \pi^+X$. The 
experiment~\cite{qi07} took place in Hall~A at the Jefferson 
Lab using a 5.09~GeV electron beam incident on a 15~cm liquid 
hydrogen target. Scattered electrons were detected in one of 
the HRS's in coincidence with electroproduced hadrons in the 
second HRS. Each spectrometer was positioned at $12.5^\circ$ 
relative to the beamline, but the use of additional septum 
magnets~\cite{septum} allowed to reach smaller production 
angles, down to $\sim6^\circ$. In this configuration, the 
spectrometers have an effective acceptance of approximately 
4~msr in solid angle and $\pm$4.5\% in momentum, while still 
maintaining their nominal $10^{-4}$ full-width half-max 
momentum resolution~\cite{HRS}. To obtain the desired MM 
coverage, the central momentum of the electron HRS was 
varied between 1.85 and 2.00~GeV, while the central 
momentum of the hadron HRS was changed between 1.89 and 
2.10~GeV. In such configuration, the average momentum transfer 
of the virtual photon was \mbox{$<Q^2>~\approx~0.1$~(GeV/c)$^2$}, 
and the average cm photon energy was 
$<E^{\text{cm}}_{\gamma^*}>~=~1.1$~GeV which means 
that $<$~W~$>$~=~2.53~GeV. For the kaon kinematics, the cm
scattering angle was 
5.6$^{\circ}\le\theta^{\text{cm}}_{\gamma^*K}\le11.4^\circ$, 
and the angular acceptance was
$\Delta\Omega^{\text{cm}}_{\gamma^*K}\approx 38$~msr.

Calibration of HRS's was based on precise measurements of the
known MM peaks for the neutron (in $\pi^+X$) and for the 
hyperons $\Lambda$(1116), $\Sigma^0(1193)$ (both in $K^+X$). 
These three baryons decay through weak or electromagnetic 
interactions, and the proper widths of their peaks are 
negligible. Therefore, the observed widths of the peaks 
directly determine the MM resolutions at the corresponding 
momenta of the registered meson ($\pi^+$ or $K^+$). Then, 
the resolution of HRS for the $\Lambda(1520)$ may be 
determined by extrapolation of those resolution values to 
the MM region of 1520~MeV. Such a procedute gave the resolution 
$\sigma=1.5~\textrm{MeV}$~\cite{qi07}. We apply this resolution 
when fitting the data.  Further experimental details can be 
found in Refs.~\cite{qi07,qi07T,qi10}.

\section{Present analysis of the $\Lambda(1520)$ }

The main initial goal of the Hall~A experiment~\cite{qi07} was
to search for possible very narrow resonances with $\Gamma\sim
1$~MeV. Therefore, the MM spectra were scanned with 1~MeV steps.
This is not adequate for the much broader resonances. To avoid 
strong fluctuations and excessively large statistical errors 
for every experimental point, we use now the 4~MeV bin size to 
analyze the data for $\Lambda(1520)$. Such a size is adequate 
for the resolution with $\sigma=1.5$~MeV, it provides smaller
statistical errors and fluctuations. 

Generally, we describe the MM spectra in the form
\begin{equation}
	{\rm Fit} = BW + BG \,, \label{fit}
\end{equation}
where $BW$ is the Breit-Wigner (BW) contribution for the resonance
$\Lambda(1520)$, and the term $BG$ combines all other contributions
(including other possible resonances), which provide a
background for the $\Lambda(1520)$.

The BW contribution may be written as
\begin{equation}
	BW = A_{BW}~\Gamma(M_X)~D(M_X)\,. \label{res}
\end{equation}
In the non-relativistic form, we have
\begin{equation}
	D^{-1}_{nrel}(M_X)=(M_X - M_0)^2
       + \Gamma^2(M_X)/4\, .
\label{nrel}
\end{equation}
In the relativistic form, this denominator should
look like $|M_X^2 -[M_0 - i\Gamma(M_X)/2]^2\,|^2 $,
but usually  $\Gamma^2\ll M^2_0$, and $D^{-1}$
may be approximately written in the form
\begin{equation}
	D^{-1}_{rel}(M_X)= (M_X^2-M_0^2)^2 + M_0^2\, \Gamma^2(M_X)\,.
\label{rel}
\end{equation}
With the energy-independent width, we just define $\Gamma(M_X)
\equiv\Gamma_0\,$. Parameters $M_0$ and $\Gamma_0$ are the 
conventional BW mass and width of the $\Lambda(1520)$. 

The energy-dependent width has a more complicated structure.
Let us recall that the total width is the sum of partial ones 
for all decay modes, $\Gamma(M_X)=\sum_i \Gamma_i(M_X)\,$. 
Here we emphasize that every partial width should have its 
own energy dependence, corresponding to the threshold and 
kinematical properties of the particular decay channel.

Decays of the $\Lambda(1520)$ are dominated by only three channels,
$\overline{K}N,\,\pi\Sigma,$ and $\pi\pi\Lambda\,$. Branching
ratios for two other modes, $\pi\pi\Sigma$ and $\Lambda\gamma$,
are not more than 1\% each~\cite{PDG}, and we neglect them here.
Also neglected is the channel $\Sigma^0\gamma$, with an even lower
branching ratio~\cite{PDG}. Further, we define $\Gamma_0=\Gamma(M_0)$.
Here we also consider $M_0$ and $\Gamma_0$ as the BW mass and width 
of the $\Lambda(1520)$. In all cases $M_0$ and $\Gamma_0$ are 
fitting parameters.

Now, for extracting the mass and width of the $\Lambda(1520)$, 
we restrict ourselves to the mass interval from 1.45 to 1.65~GeV 
(this corresponds to 13,070 detected events). To the peak, we 
apply the BW term (\ref{res}), while the background is
described by various combinations.
Further, for the best-fit procedure, we use two not 
quite equivalent methods, least-squares (min-$\chi^2$) and 
log-likelihood (LL) ones. In such a way, we obtain several sets 
of numerical values for pairs ($M_0,\,\Gamma_0$) and can trace 
their dependence on the assumptions used.

\section{Results and their Discussions}

We begin with considering changes of the BW mass $M_0$. They
appear to be rather small, though not always negligible. The
non-relativistic expression (\ref{nrel}) for the BW term provides
$M_0$ values lower than the relativistic Eq.(\ref{rel}), at the
level of $\sim0.05$~MeV.  Similarly, the LL procedure gives lower 
$M_0$ than min-$\chi^2$; the difference may be as small as 0.01~MeV, 
but may reach $\,\sim0.15$~MeV. Structure of the background can 
also shift $M_0\,$; the difference for various variants which we 
use is, again, not more than 0.15~MeV.

A more essential effect comes from energy dependence of the width.
Description with the energy-dependent width results in $M_0$ about
0.6~MeV lower than for the energy-independent width. Interestingly,
it is at the same level as our statistical uncertainty for $M_0$,
which is \mbox{$\sim0.6$~MeV} in all the studied cases.

Changes of the BW width $\Gamma_0$ also show some regularities.
Shifts of results for using the (non-)relativistic Eqs.(\ref{nrel})
or (\ref{rel}) are not more than 0.06~MeV. The LL fitting gives a
lower width than min-$\chi^2$; the difference may be 0.2~MeV, but
may reach $\sim0.9$~MeV (for comparison, our statistical
uncertainty for the width is $\sim2$~MeV).

More complicated is the influence of the background description.
Most influential in fitting our data is contribution of the 
resonance $\Lambda(1405)$. Its exclusion diminishes $\Gamma_0$; 
the difference may be up to $\sim1$~MeV (the corresponding shift 
of $M_0$ is much smaller, not more than 0.1~MeV).

Now we are able to formulate some conclusions, which may have a 
more general character.
\begin{itemize}

\item{The width of $\Lambda(1520)$ is sufficiently small, so the
relativistic and non-relativistic forms give practically the same
values of $M_0$ and $\Gamma_0$ (at the present level of accuracy).}

\item{By definition, the LL fitting always provides a larger value
of $\chi^2\,$, than the min-$\chi^2$ fitting. However, formally
they should be equivalent at asymptotically high statistics. In
this sense, the present statistics is not asymptotical yet. Both
$M_0$ and $\Gamma_0$ are different in the two fittings, the
differences are comparable to the statistical uncertainties of
those BW parameters. In terms of $\chi^2$ per degree of freedom,
$\chi^2$/dof, which is typically $\,\sim1.5$ in our studies here,
the LL fitting is up to 0.05 higher than min-$\chi^2\,$.}

\item{The $M_0$ has a smaller statistical uncertainty and is less
affected by any change of fitting procedure than the $\Gamma_0$.
The width is especially sensitive to the background form.}

\end{itemize}

\begin{figure}[th]
\centerline{
\includegraphics[height=0.35\textwidth, angle=0]{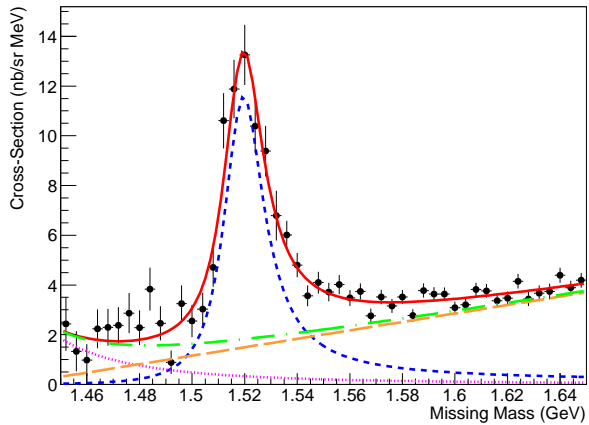}}
\vspace{3mm} \caption{(Color on line) Fit to the experimental MM 
	distribution is shown by the solid line. The short-dashed 
	line is the BW contribution for the $\Lambda(1520)$ (note 
	an asymmetric form, due to the energy-dependent width). 
	The total background (dot-dashed line) is the sum of the 
	linear binomial (long-dashed line) and the BW tail of the 
	$\Lambda(1405)$ (dotted line). \label{fig:g3}}
\end{figure}

For extracting our resulting BW parameters, we use the BW
relativistic expression (\ref{res}),\,(\ref{rel}) with
energy-dependent width (both points are theoretically
motivated to be more reasonable). 
The
corresponding least-squares (min-$\chi^2$) fit in the mass
interval 1.45 - 1.65~GeV is shown in Fig.~\ref{fig:g3}, together
with all separate contributions. This fit corresponds to
$\chi^2$/dof=1.46 and gives
\begin{equation}
	M_0 = 1520.4\pm 0.6~{\rm MeV}\,,~~ \Gamma_0 = 18.6\pm 1.9~{\rm
	MeV} \,, \label{MG}
\end{equation}
with pure statistical uncertainties. Each of the two BW
parameters has also a systematic uncertainty. It is about 
1.5~MeV, mainly due to the mass scale uncertainty, for $M_0$, 
and about 1~MeV, mainly due to the fitting procedure
ambiguity, for $\Gamma_0$.

For comparison, the same description with the LL fit gives
slightly lower values
\begin{equation}
	M_0 = 1520.3\pm 0.6~{\rm MeV}\,,~~ \Gamma_0 = 17.8\pm 1.9~{\rm
	MeV} \,, \label{MGL}
\end{equation}
and the higher value $\chi^2$/dof=1.50 . 

Even without accounting for the systematic uncertainties, both 
$M_0$ and $\Gamma_0$ are in reasonable agreement with previous
experimental values and with their RPP average~\cite{PDG}. It is 
worth to note, however, that the uncertainties in the later works
used in RPP are larger than those in the earlier works. This may 
hint that the uncertainties stated in the earlier works (and, 
therefore, in the average) are too optimistic.

Now we can discuss position of the $S$-matrix pole corresponding
to the $\Lambda(1520)$, that has never been discussed for hyperons. 
It is determined by vanishing of the denominator (\ref{rel}), 
\textit{i.e.}, by a solution of the equation
\begin{equation}
	(W_p^2-M_0^2)^2 + M_0^2\, \Gamma^2(W_p)=0\,. \label{pole}
\end{equation}
This non-linear equation may have non-unique solutions. Physically
reasonable is only one of them, close to $(M_0-i\, \Gamma_0/2)\,$.
It is convenient, therefore, to rewrite Eq.(\ref{pole}) in the
form
\begin{equation}
	W_p =M_0 \left[1- i\,\Gamma(W_p)/M_0\,\right]^{1/2}\,. \label{pol}
\end{equation}
Its complex solution gives the pole mass $M_p= \textrm{Re}\,W_p$
and the pole width $\Gamma_p=-2\,\textrm{Im}\,W_p$.

%
Numerical solution with values (\ref{MG}) for $M_0$ and $\Gamma_0$ 
gives
\begin{equation}
	M_p=1518.8~\textrm{MeV} \,,~~~ \Gamma_p=17.2~\textrm{MeV}\,.
\label{MpGp}
\end{equation}
Note that $M_p<M_{BW}$, $\Gamma_p<\Gamma_{BW}$, with the mass
difference exceeding the statistical uncertainty. Such relation
for masses may be rather general, as suggested by comparison 
with the mass pairs (BW and pole) shown for $\pi N$
resonances in Listings of RPP~\cite{PDG}. This assumption 
is supported by the analysis of Eq.(\ref{pol}) presented 
in Ref.~\cite{qi10}.

%

In summary, we have found the mass and width of the 
$\Lambda(1520)$ in near-forward electroproduction. The 
extracted BW parameters of the resonance are shown to depend 
not only on experimental data, but also on the way of their 
treatment. The extracted width is more sensitive to the various 
treatments than the mass. For the $\Lambda(1520)$, the 
non-resonance background should be accurately studied and 
understood if one intends to extract the mass and, 
especially, width with uncertainty of order 1 - 2~MeV. 
Having the BW mass and width (\ref{MG}), we also give the 
first estimate (\ref{MpGp}) of the pole parameters for the 
$\Lambda(1520)$.  The pole values for both mass and width 
tend to be lower than the BW values.

\section{Acknowledgments}

The authors express their gratitude to R.~Arndt and to E.~Pasyuk
for useful discussions. This work
was supported in part by the U.~S.~Department of Energy under
Grants DE-FG02-99ER41110 and DE-FG02-03ER41231, by the Russian
State grant SS-3628.2008.2, and by Jefferson Science Associates
which operates the Jefferson Lab under DOE contract DE-AC05-06OR23177.



\begin{thebibliography}{99}
\bibitem{PDG}   K.~Nakamura \textit{et al.} (Particle Data Group),
                J.\ Phys.\ G\ \textbf{37}, 075021 (2010).
\bibitem{oset}  L.S.~Geng, E.~Oset, and B.S.~Zou, Eur.\ Phys.\ J.\
                \textbf{A}\ \textbf{38}, 239 (2008).
\bibitem{zou}   J.J.~Wu, S.~Dulat, and B.S.~Zou, Phys.\ Rev.\ 
		C\ \textbf{81}, 045210 (2010).
\bibitem{qi10}  Y.~Qiang \textit{et al.} 
                arXiv:1003.5612 [hep-ph].
\bibitem{HRS}   J.~Alcorn \textit{et al.}, Nucl.\ Instrum.\ Methods\
                \textbf{A}\ \textbf{522}, 294 (2004).
\bibitem{qi07}  Y.~Qiang \textit{et al.} (JLab Hall~A Collab.),
                Phys.\ Rev.\ \textbf{C}\ \textbf{75}, 055208 (2007).
\bibitem{septum}P.~Brindza \textit{et al.}, IEEE Trans\ Appl.\
                Supercond.\ \textbf{11}, 1594 (2001).
\bibitem{qi07T} Y.~Qiang, PhD Thesis, MIT, 2007.

\end{thebibliography}
\end{document}